\documentclass{IEEEtran}
\usepackage{cite}
\usepackage{amsmath,amssymb,amsfonts}
\usepackage{graphicx}
\usepackage{textcomp,nicefrac}

\usepackage{color}
\usepackage{braket}
\usepackage{subfigure}
\usepackage{siunitx}

\def\BibTeX{{\rm B\kern-.05em{\sc i\kern-.025em b}\kern-.08em
T\kern-.1667em\lower.7ex\hbox{E}\kern-.125emX}}
\begin{document}
\title{SQUID G.A.M.E.: Gamma, Atmospheric, and Mono-Energetic Neutron Effects \\on Quantum Devices}
\author{Gioele Casagranda, Elizabeth Auden, Carlo Cazzaniga, Maria Kastriotou, Christopher Frost, Marzio Vallero, Flavio Vella,  and Paolo Rech
\thanks{This work was supported partly by the INFN section 5 project QuRE, by the Q@TN lab, and by the Italian Ministry for University and Research (MUR) through the Departments of Excellence 2023-27 program under Grant L.232/2016 awarded to the Department of Industrial Engineering. Beam time at ChipIR/NILE was awarded thanks to UKRI ISIS Neutron and Muon Source.
}
\thanks{Gioele Casagranda, Marzio Vallero, and Flavio Vella are with the Department of Information Engineering and Computer Science of the University of Trento, Italy (e-mail: gioele.casagranda@unitn.it, marzio.vallero@unitn.it, flavio.vella@unitn.it).}
\thanks{Elizabeth Auden is with the Los Alamos National Laboratory, USA LA-UR-25-26629 (e-mail: auden@lanl.gov)}
\thanks{Carlo Cazzaniga, Maria Kastriotou, and Christopher Frost are with the Science and Technology Facility Council, UK (e-mail: christopher.frost@stfc.ac.uk, maria.kastriotou@stfc.ac.uk,
carlo.cazzaniga@stfc.ac.uk}
\thanks{Paolo Rech is with the Department of Industrial Engineering of the University of Trento, Italy (e-mail: paolo.rech@unitn.it).}
\thanks{© 2025 IEEE. Personal use of this material is permitted. Permission from IEEE must be obtained for all other uses, in any current or future media, including reprinting/republishing this material for advertising or promotional purposes, creating new collective works, for resale or redistribution to servers or lists, or reuse of any copyrighted component of this work in other works.}}

\maketitle

\begin{abstract}
Quantum devices are a promising solution to many research applications, including medical imaging, precision magnetic field measurements, condensed matter physics, and overcoming the limits of classical computing. Among the available implementations, the superconducting technology is the current focus of scientific research and industrial applications, excelling in performance and scalability. Despite this, superconducting quantum systems are extremely prone to decoherence, and in particular, they are highly sensitive to radiation events.

In this paper, we analyze the response of a superconducting device (SQUID) to radiation. We expose the SQUID to beams of monoenergetic 14 MeV neutrons (NILE - ISIS), atmospheric~1-800 MeV neutrons (ChipIR - ISIS), and gamma rays with 1.25 MeV average energy (CALLIOPE - ENEA).

These experiments show that the SQUID is sensitive to the two neutron fields, while gamma rays at 1.25 MeV leave it mostly unaffected. Following our experiments with neutrons, it is possible to characterize the SQUID's response and even classify faults according to their shape and duration. We identify two categories: bursts (long lasting) and peaks (short lived).
To investigate the different responses to neutrons and gamma rays, we employ Geant4 simulations, which highlight differences in the deposition spectra and the energy propagation, but likewise predict the vulnerability of the SQUID in both cases.
\end{abstract}


\section{Introduction}
\label{sec:introduction}

\IEEEPARstart{Q}{uantum} technologies have established themselves as a compelling research field and are emerging as highly requested commercial solutions. There exists a multitude of different approaches to exploit the quantum properties of matter; superconducting-based quantum devices, such as Superconducting Quantum Interference Devices (SQUIDs), are among the most popular and promising. Indeed, they offer easy and well-established fabrication procedures, great scalability, and fast and efficient manipulation of the encoded information. However, this generally comes at the cost of a very high susceptibility to intrinsic noise and radiation fields. While the former is already being addressed with shielding, active noise control circuits, and algorithm solutions.~\cite{Andersen2020,  Tiurev2023correcting, Katsuda2024, vallero2024efficacy}, the latter is yet to be fully understood, and it is becoming a bottleneck for the exploitation of quantum technologies. 

Some preliminary studies have shown, through field tests~\cite{mcewen2022resolving}, GEANT4 simulations~\cite{simulazioni}, and neutron experiments~\cite{elizabeth_nsrec23} that (i) quantum devices exhibit a high sensitivity to radiation, orders of magnitude higher than CMOS transistors, (ii) the persistence of transient faults can be in the order of $\sim \SI{100}{\mu s}$, and (iii) the spread of energy in the substrate affects qubits even several millimeters away from the particle impact. Previous works, while showcasing the urgency of addressing quantum devices' radiation reliability~\cite{nature_rad, Martinis2021,muons2021, Cardani_2023}, still lack a characterization of radiation impact on the device's operative parameters. Moreover, the exact correlations between the observed effects and impinging particles are still largely unknown.



In this paper, we aim to advance the knowledge on quantum devices' reliability by studying the radiation response of a SQUID~\cite{SquidHandbook_introduction}, the most fundamental superconducting quantum device, and the building block of more highly sophisticated quantum systems. We test the device with three different radiation fields. At ISIS Muon and Neutron Source, we test with monoenergetic neutrons at \SI{14}{MeV} (NILE) and with atmospheric-like neutrons in the range \SI{1}{}-\SI{800}{MeV} (ChipIR). At ENEA Casaccia, we test with a source of $^{60}$Co emitting gamma rays with an average energy of \SI{1.25}{MeV} (CALLIOPE).
Considering the pioneering attitude of the experiment, our contribution is threefold. (i) Our first result is the verification of the suitability of NILE, ChipIR, and CALLIOPE facilities for quantum device experiments. (ii) In second place, we provide a punctual characterization of the response of the SQUID under test to different radiation fields. (iii) We also propose a deterministic separation between radiation-induced and non-radiation-induced events, categorizing the former into two classes. We identify \textit{peak-type} and \textit{burst-type} faults, correlating their characteristics with fundamental physics principles.

The interpretation of the experimental data is supported by a deep analysis employing Monte Carlo simulations in Geant4~\cite{Geant4} and G4CMP~\cite{g4cmp}. This complementary research accounts for calculating and describing the energy deposition and propagation in the system. Ultimately, thanks to the simulations, it is possible to go beyond the characterization of the SQUID response, giving an interpretation to the phenomena that drive it.

The spirit of this work is challenging, and most of the tests are not straightforward in their design and realization. From the first tests at NILE and ChipIR we fine-characterize the SQUID response and classify faults according to their features. On the contrary, the test with gamma rays unexpectedly did not provide observable events and the received dose is likely to cause permanent (or at least long-term) damage to the SQUID, rendering the subsequent tests at NILE and ChipIR inconclusive.

The remainder of the paper is structured as follows. Section~\ref{sec:background} serves as a background on the functioning of superconducting quantum devices. In Section~\ref{sec:experimentalMethodology}, we present our experimental setup and methodology. Section~\ref{sec:simulations} shows simulation modelling of the experiments, to give insights into the expected results. Section~\ref{sec:experimentalResults} highlights the results for the response of the SQUID and the outcomes for energy deposition and propagation obtained through the simulations.
At last, in Section~\ref{sec:conclusions} we provide a discussion on the main findings, with an analysis on the experimental limitations and future investigation directions.

\section{Background}
\label{sec:background}
\begin{figure}[t]
    \centering
    \includegraphics[width=\columnwidth]{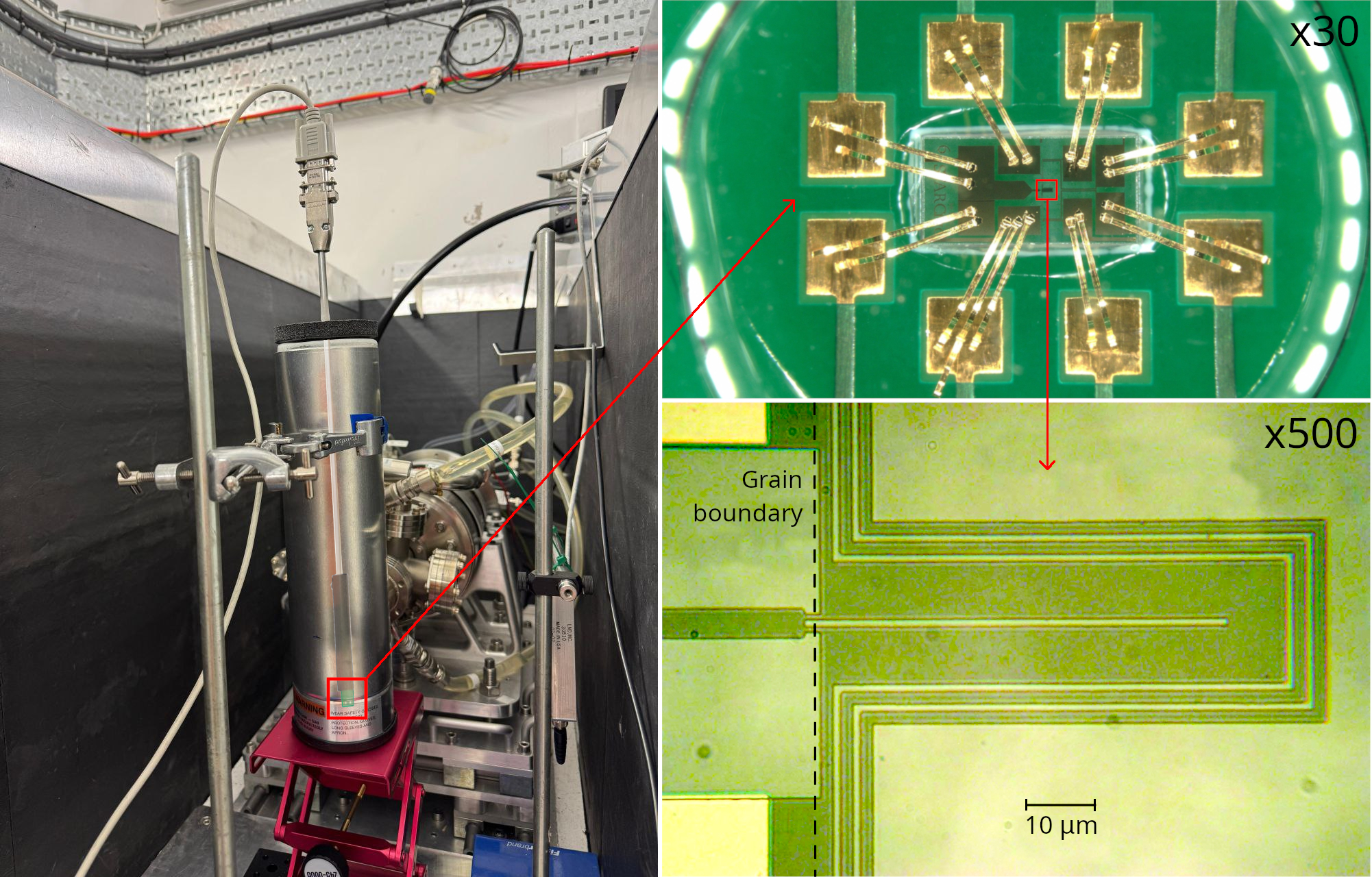}
    \caption{Experimental setup at NILE and closeup of Mr. Squid. The squid device is superimposed in transparency to show its orientation inside the dewar (left). The microscope view of the Squid device at x30 magnification (top right, from \cite{elizabeth_nsrec23}) and at 500x magnification (bottom right). The darker part is the photolithographed YBCO layer.} 
    \label{fig_setup_NILE}
\end{figure}





The radiation testing of quantum devices poses significant challenges. In the first place, the experiment follows quantum physics principles, but the measured output is classical. In the hereinafter, we will provide a concise, albeit complete, explanation of the quantum phenomena involved in our experiment and an intuition on how our measures are a witness to a reliability problem.

\subsection{Superconducting Quantum Devices Essentials}
Not only can quantum devices such as SQUIDs measure magnetic fields with high precision, they can also be used to encode a superposition of states. Any arbitrary quantum state is indeed represented mathematically as the linear combination of two different states: $\ket{\Psi}=\alpha \ket{0}+\beta \ket{1} $, where $\{\ket{0},\ket{1}\}$ are the states, and $\alpha$, $\beta$ are complex numbers that represent the respective amplitudes, with the constraint that $1=\alpha^2+\beta^2$~\cite{Nielsen_Chuang_2010}.
Following the laws of physics, the only way to retrieve a quantum mechanical state is to let its respective wavefunction collapse in a controlled environment (measurement), iterating this procedure multiple times in order to infer the amplitude values. A complete explanation of a quantum measurement is beyond the scope of this work, and it is not necessary for the sake of our tests, which rely purely on classical measurements.

\begin{figure*}[!th]
    \centering
    \includegraphics[width=\linewidth]{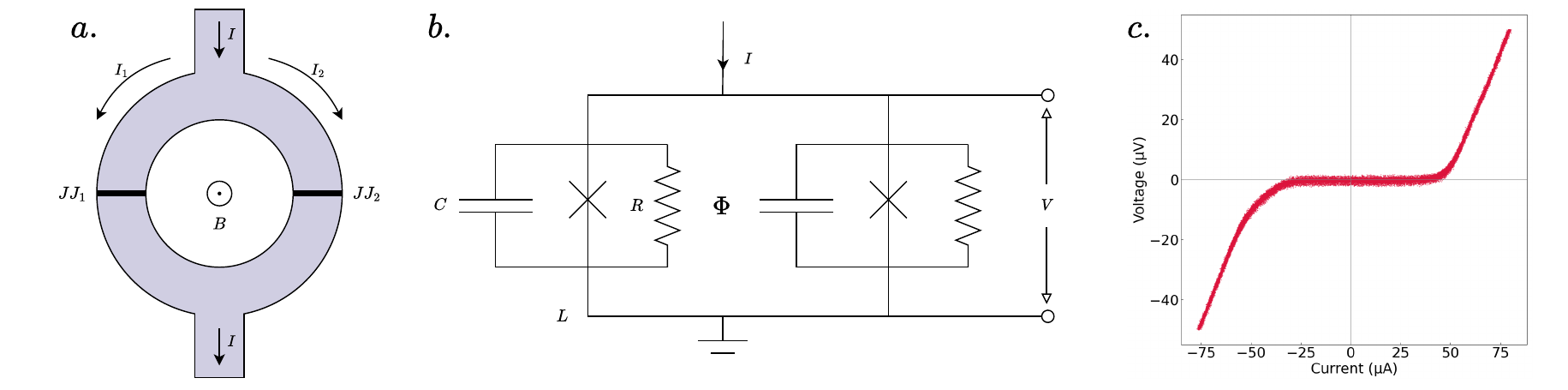}
    \caption{Circuit description and characteristic of the Mr. Squid device. a. The SQUID loop schematic representation: in gray the superconductor and in black the two Josephson junctions. The supercurrent \textit{I} and the magnetic field \textit{B} are indicated, too. b. The electronic scheme of the SQUID. The two Josephson junctions are represented by a capacitor, a resistance, and a nonlinear inductance in parallel. c. The current-voltage characteristic behavior of the device. The measured critical current is $I_c=54.3 \pm 0.7 \text{$\mu$A}$ and for $|I|< I_c$ the output voltage is null. For greater values, the current and voltage follow the standard ohmic relation.}
    \label{squid_descriptor}
\end{figure*}

Quantum states are physically encoded into 2-level quantum systems, which can be built by exploiting different technologies: atomic orbitals, trapped ions, and, most commonly, superconducting circuits. The discussion in this paper is focused on the latter technology, where the 2 energy levels are obtained with a resonant non-linear LC circuit. By employing a superconducting circuit and cooling it below the conductor's critical temperature, it is possible to build a harmonic potential with equally spaced energy levels, and theoretically no energy loss. In order to get access to two well-defined energy levels, 
the circuit is integrated with two nonlinear inductances, called Josephson Junctions (JJs), to introduce an anharmonicity in the system, which in turn exposes two addressable energy levels. This circuit is known as SQUID loop, or just SQUID (Superconducting QUantum Interference Device) \cite{roth2021introduction}.

The SQUID functionality is then the combination of the superconductivity properties of the metal utilised and the microscopic behavior of the Josephson Junctions~\cite{SquidHandbook_introduction,SquidHandbook_theory,SquidHandbook_DC}. The superconducting regime allows for a resistive-less flow of Cooper pairs current through the circuit, which ultimately means a stable encoding not subject to energy loss. The Josephson Junction brings the needed non-linearity by introducing an ultra-thin energy barrier in the superconductor, which can be crossed by Cooper pairs via the tunnelling effect.

While in the superconducting regime, the V-I characteristic of a SQUID shows a null voltage output in response to a non-null driving current. This behavior is maintained lossless while the driven current is kept under the critical value $I_c$.

\subsection{The Reliability Problem}

One of the strengths of superconducting quantum devices over other technologies is represented by the ease of implementation of communication channels. This inherently brings to the picture accidental interactions, too. Quantum states implemented using superconducting quantum devices are extremely prone to decoherence by intrinsic or external agents, namely radiation.
The effects of radiation on quantum devices have been studied through on-field experiments by multiple teams, both considering various quantum systems \cite{Cardani2021, nature_rad,mcewen2022resolving}. These works agree on the seriousness of the problem and identify three main strands of evidence:  (a) even light particles can deposit enough energy to affect quantum devices' operation; (b) fault persistence can be long, in relative terms, up to hundreds of seconds; (c) superconducting quantum devices placed on a common substrate can be affected simultaneously by a single particle strike.

The literature presents some gaps in the systematic study of the microscopic behavior of superconducting quantum devices when affected by radiation. Thanks to GEANT4~\cite{Geant4} and G4CMP~\cite{g4cmp}, it is possible to study the phenomenon at a simulation level, and indeed both characterization~\cite{simulazioni} and mitigation strategies~\cite{yelton2024} can be found in the literature. Notably, however, a thorough experimental characterization of the most fundamental superconducting quantum device (the SQUID) is still missing. This research direction can give important insight to the community to systematically face the reliability issue, a problem that is currently only addressed with non-scalable solutions such as caves and shields~\cite{dedominicis2024evaluatingradiationimpacttransmon, Cardani2021, underground1}.

\subsection{Radiation-induced fault mechanism}
To tailor the radiation reliability problem to the SQUID under test, it is crucial to investigate the behavior of superconductivity and the Josephson effect under radiation exposure. Superconductors can certainly be affected by radiation. Indeed, irradiating these components with neutrons and gamma rays does alter both their macroscopic properties, such as the material's critical temperature, as well as their microscopic lattice \cite{Nicholls2022, Zheng_2024}. Such problems are profusely studied for scopes generally most unrelated to quantum technologies. We infer that, although these mechanisms may play a role in affecting the performance or the durability of quantum devices, they are most definitely not responsible for the aforementioned transient effects. Conversely, since the state encoding depends on the flow of Cooper pairs, whose binding energy is in the order of meV, it is manifestly evident that SEUs (Single Event Upsets) in quantum devices are induced by an \textit{alteration} of the Josephson effect.

\section{Experimental Methodology}
\label{sec:experimentalMethodology}

In this section, we introduce the quantum device under test, present the three radiation facilities, and describe the methodology designed to identify radiation-induced events.

\subsection{SQUID and test setup}
\label{subsec_SQUID}

In Section~\ref{sec:background}, we introduced the functioning of the SQUID on a theoretical level, but its working principle highly depends on the specific structural characteristics. The SQUID's original purpose is to measure magnetic fields with extreme precision; therefore, it is highly sensitive to external electromagnetic fields. We are interested in characterizing how radiation affects the SQUID in a static environment, in the absence of varying magnetic fields. Doing so poses non-trivial challenges to the design and implementation of the experimental setup. 

We test the Mr. SQUID by STAR Cryoelectronics, LLC~\cite{MrSQUID_userGuide} 
(Figure \ref{fig_setup_NILE}), made of a strontium titanate (SrTiO$_3$, which is commonly named STO) substrate, on top of which a metallic layer is placed through photolithography. The latter is fabricated in the shape of a loop using yttrium barium copper oxide (Y$_1$Ba$_2$Cu$_3$O$_7$, commonly referred to as YBCO). The photolithographic process is interrupted at two points of the loop to build two grain boundary Josephson junctions. The STO substrate is \SI{1.2}{} $\times$ \SI{2.6}{} $\times$ \SI{1}{mm} large.
The superconducting layer, $\SI{25}{}$ $\times$ $\SI{150}{\mu m}$, is roughly centered on the substrate, and encircled by other metallic plates of similar dimension (Figure~\ref{fig_setup_NILE}).

As explained in Section~\ref{sec:background}, the SQUID must be operated in the superconducting regime; to do so, YBCO needs to be cooled under its critical temperature $T_c=\SI{90}{K}$. Given the YBCO high $T_c$, the superconducting regime can be easily achieved by employing liquid nitrogen ($T=\SI{77}{K}$) as coolant. STAR Cryoelectronics bundles Mr. SQUID together with a glass dewar ($d=\SI{8.5}{cm}$), shielded by a thin layer of aluminum, and a polystyrene cap that helps isolate the system for more than 15 hours. The SQUID is connected to a commercial-grade control device, provided by the same company, via a serial cable $\SI{3}{}-\SI{7}{m}$ long. The control device sets the input/output channels of the SQUID and lets the user tune the current flow through the JJs and the magnetic flux concatenated to the loop. The output signal is transmitted through two coaxial cables ($\sim \SI{1}{}-\SI{35}{m}$) to an oscilloscope, which in our setup is a Picoscope 5000 series with $ \SI{200}{MHz}$ bandwidth. The oscilloscope is connected to a laptop with a USB connector, and the laptop performs an automatic reading of the current and voltage lines to detect radiation-induced alterations.

\begin{figure}[t]
    \centering
    \includegraphics[width=1\columnwidth]{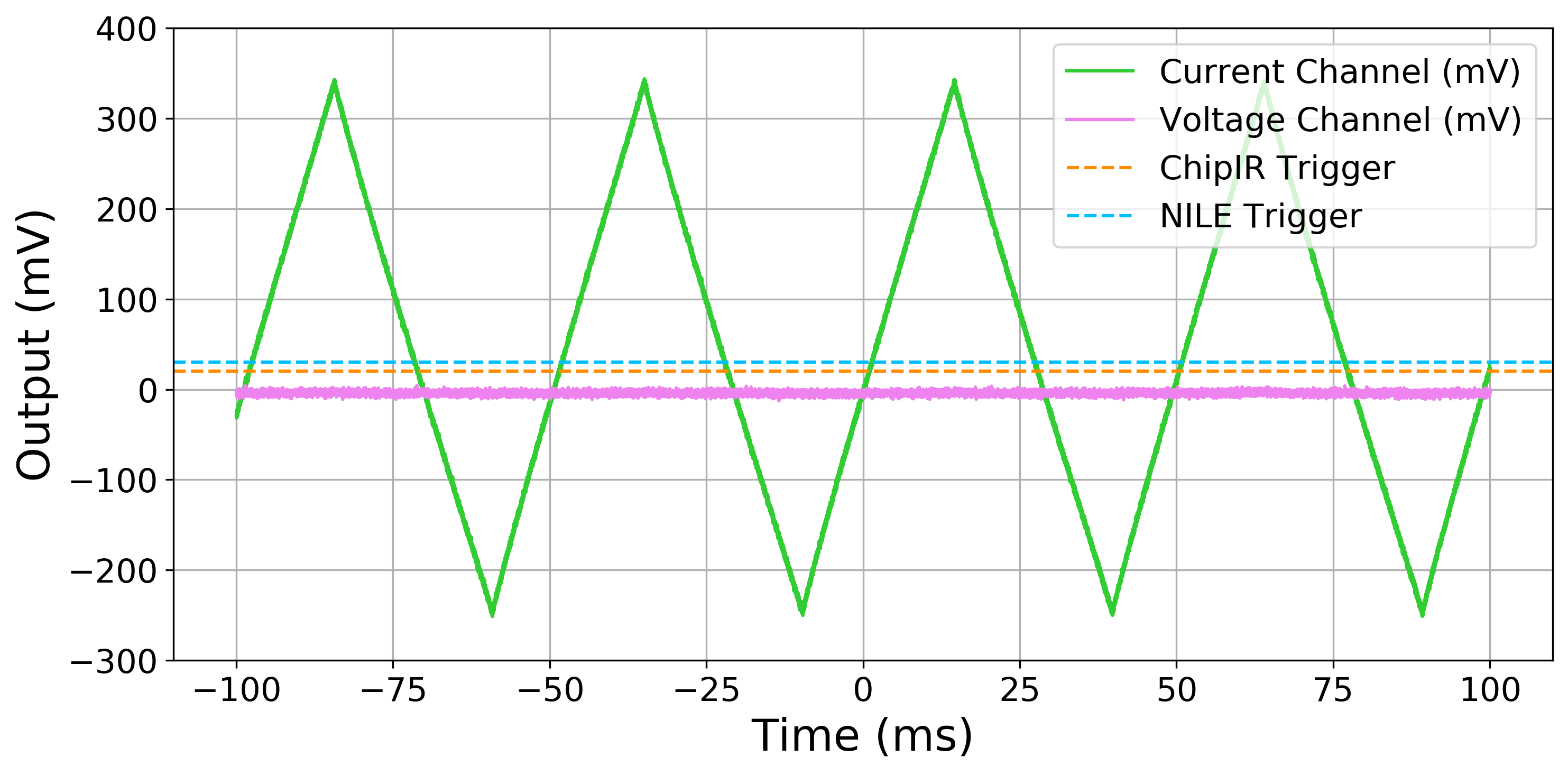}
    \caption{Current (green) driven through the SQUID with a triangle-shaped 20 kHz current waveform. In pink, the characteristic null output (in the unperturbed regime), since the current never exceeds the critical value.
    The dotted lines represent the trigger for NILE (light blue) and ChipIR (orange) during E1 and E2, respectively. On the vertical axis, the output is reported as read on the oscilloscope, so it goes through the SQUID electronics amplification. To retrieve the actual current and voltage values, it is needed to divide by a factor $\SI{e4}{}$, with the appropriate unit (- and $\Omega$)}
    \label{fig_normal_output}
\end{figure}

Before the start of every experiment, we set the ground signal by regulating the concatenated magnetic flux such that the SQUID reaches its maximal $I_{c}$ value. We then drive the device with a triangle-shaped $\SI{20}{kHz}$ current waveform with an amplitude smaller than the critical value (Figure~\ref{fig_normal_output}). This sets the SQUID in the superconducting regime, such that any non-null voltage signal must be due to a fault (either radiation- or setup-induced).

\subsection{Radiation Facilities}
\label{subsec_radiationfacilities}
Our research spans over three different radiation facilities and is a collection of five correlated experiments.

The first facility considered is NILE, within the ISIS Muon and Neutron Source research centre at the Rutherford Appleton Laboratories (UK). NILE includes two compact neutron sources which make use of two isotopes of hydrogen: deuterium and tritium~\cite{nile}. By inducing the collision of an isotope of deuterium against a tritium target, NILE generates a monoenergetic isotropic neutron field at $\SI{14}{MeV}$ with a maximum emission rate $\sim 10^{10} \ \text{neutrons/s}$.

At ISIS Muon and Neutron Source, we also test at ChipIR facility. This beam line offers fast neutrons generated by the collision of $\SI{800}{MeV}$ protons with a tungsten target~\cite{ChipIR1}. The induced neutron beam is designed and optimized to resemble the atmospheric neutron spectrum in the range $\SI{1}{}-\SI{800}{MeV}$. The beam is collimated with an average section of $\SI[parse-numbers=false]{3 \times 3}{cm^2}$ during our experiments (but can be tuned to larger sizes) and the flux in the facility is $\sim \SI{5e6}{cm^{-2}s^{-1}}$.

The third facility employed is CALLIOPE at the ENEA Casaccia research centre (IT). Multiple $^{60}$Co sources, placed directly in the irradiation chamber, generate a gamma field with an average energy of $\SI{1.25}{MeV}$~\cite{calliope}. The activity of the source is~$\sim~\SI{e10}{Bq}$, and devices can be irradiated at a maximum dose rate of $\SI{7.3}{kGy/h}$.

\subsection{Experimental details}
\label{subsec_experimentdetails}

The five experiments presented in this work have been carried out over a period of six months, from December 2024 to June 2025. In detail, the first two experiments were conducted between 15/12/24 and 20/12/24 at NILE (E1) and ChipIR (E2). The third experiment was conducted between the 28/04/25 and 29/04/25 at CALLIOPE (E3). The fourth experiment was conducted at NILE from 18/05/25 to 20/05/25 (E4). Finally, the last experiment was conducted from 04/06/25 to 06/06/25 at ChipIR (E5). All the experiments involved the same SQUID probe, the same control device, the same oscilloscope, and the same recording software. Differences, pointed out hereinafter, regard forced choices of coaxial and serial cables. During the time windows between the experiments, the SQUID was mostly left isolated to let the radioactive byproducts of radiation exposure decay safely. Occasionally, however, tests in the absence of artificial irradiation were performed to check the nominal operation status of the device.

\subsubsection{E1 and E2}

At NILE, the dewar is placed $\sim~\SI{0.5}{cm}$ away from the neutron source irradiating isotropically, resulting in the SQUID being $\SI{5}{cm}$ away from the source (Figure~\ref{fig_setup_NILE}). The emission rate of neutrons is in the order of $\SI{e9}{n/s}$, so the average flux on the probe is $\bar{\phi}_n^{\text{NILE}} \simeq \SI{3.3}{} \times \SI{e6}{n/(s \cdot cm^2)}$.
During our test time at ChipIR the syncrotron is running with reduced current ($\SI{30}{\mu A}$) and consequently the neutron flux (above $\SI{10}{MeV}$) on the SQUID is $\bar{\phi}_n^{\text{ChipIR}} \simeq \SI{1.9e6}{n/(s \cdot cm^2)}$. By convention, the atmospheric flux is computed only considering neutrons above $\SI{10}{MeV}$, which means $\bar{\phi}_n^{\text{ChipIR}} \doteq \bar{\phi}_{n | E>10 MeV}^{\text{ChipIR}}$.

In the two facilities, the setups are analogous, with the only difference being the length of the serial cable connecting the SQUID to its control device. At NILE the cable is $\SI{3}{m}$ long, whereas at ChipIR it is $\sim \SI{5}{m}$. The control device, in both experiments, is placed in a protected area inside the irradiation rooms.
Two coaxial cables~$\sim~\SI{35}{m}$ are employed to connect this device to the experiment control room, where the oscilloscope and the computer running the code used for measurement are placed.

We test the SQUID for about $\SI{4.5}{hours}$ at NILE and $\SI{14}{hours}$ at ChipIR, which accounts for a fluence of about $\SI{5.3e10}{n/cm^2}$ and $\SI{9.6e10}{n/cm^2}$, respectively. The two data collections take place $\SI{20}{hours}$ apart. During the interval between the two data collections, as well as before the first data collection, the device is exposed to neutrons for short windows of time in order to tune the experimental setup.

Neutron generation and subsequent interactions with the environment also serve as an indirect source of gamma rays, which, as shown by previous 
works~\cite{simulazioni,Cardani_2023,larson2025quasiparticlepoisoningsuperconductingqubits} and presented in Section~\ref{sec:experimentalResults}, can contribute to the quantum device rate of electrical faults. Simulations and legacy measurements in the test areas allow us to compute the average gamma flux in the two facilities. Conversely to the neutron field, the abundance of gamma rays is almost independent of the SQUID position in the irradiation line. 
At ChipIR, gamma ray flux is calculated solely with simulation, and a large uncertainty is expected, due to the complexity of the field and the lack of cross sections at high neutron energy in the databases. Conversely, at NILE, the gamma ray flux is better known because the field is simpler and not pulsed, allowing for benchmark measurements, and nuclear cross sections are well known at 14 MeV. $\bar{\phi}_\gamma^{\text{NILE}} \simeq \SI{2.2e5}{\gamma/(s \cdot cm^2)}$, which is about $7\%$ of the neutron flux at the SQUID position.


\begin{figure}[t]
    \centering
    \subfigure[Sawtooth false positive]{
        \includegraphics[width=0.22\textwidth]{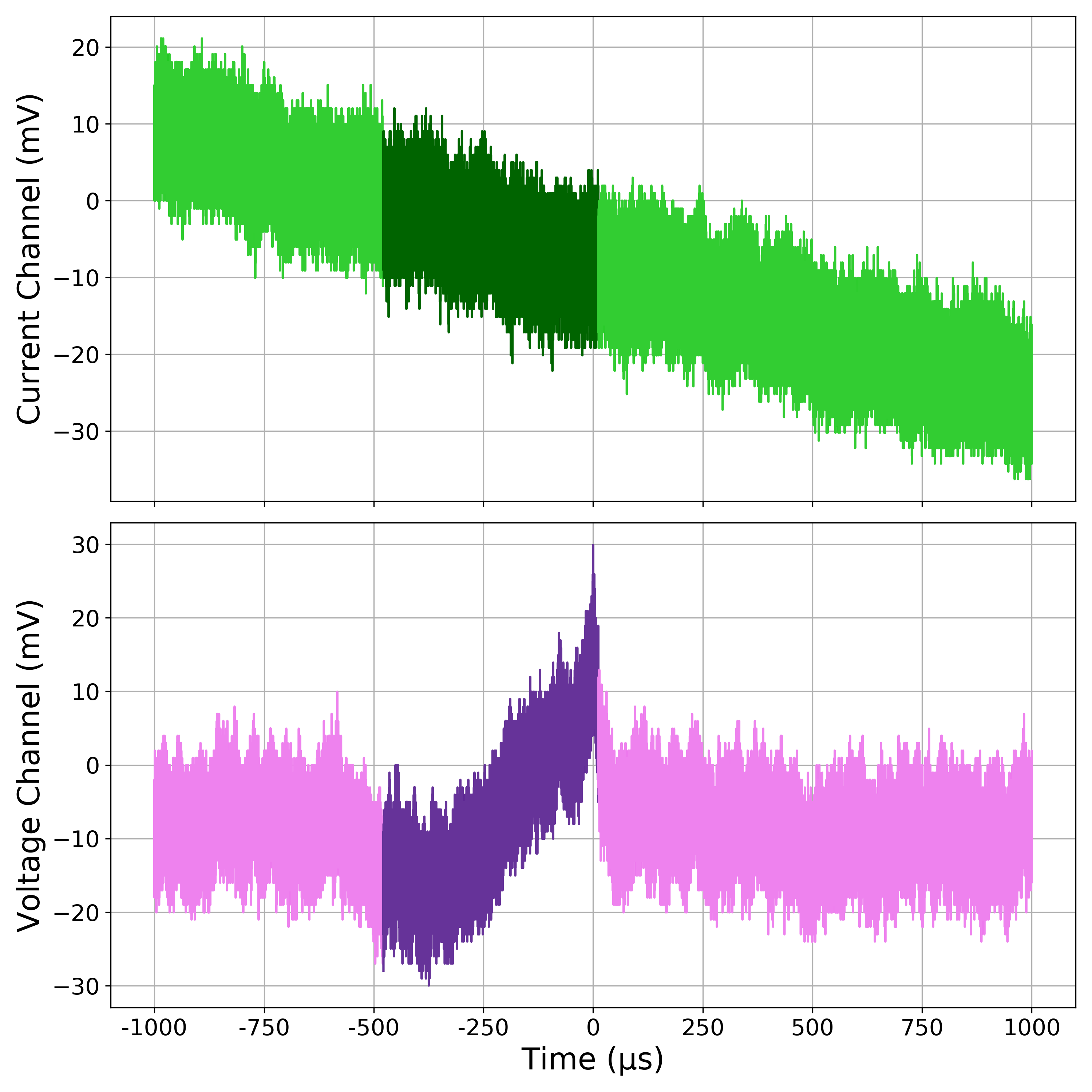}
    }
    \subfigure[Oscillating false positive]{
        \includegraphics[width=0.22\textwidth]{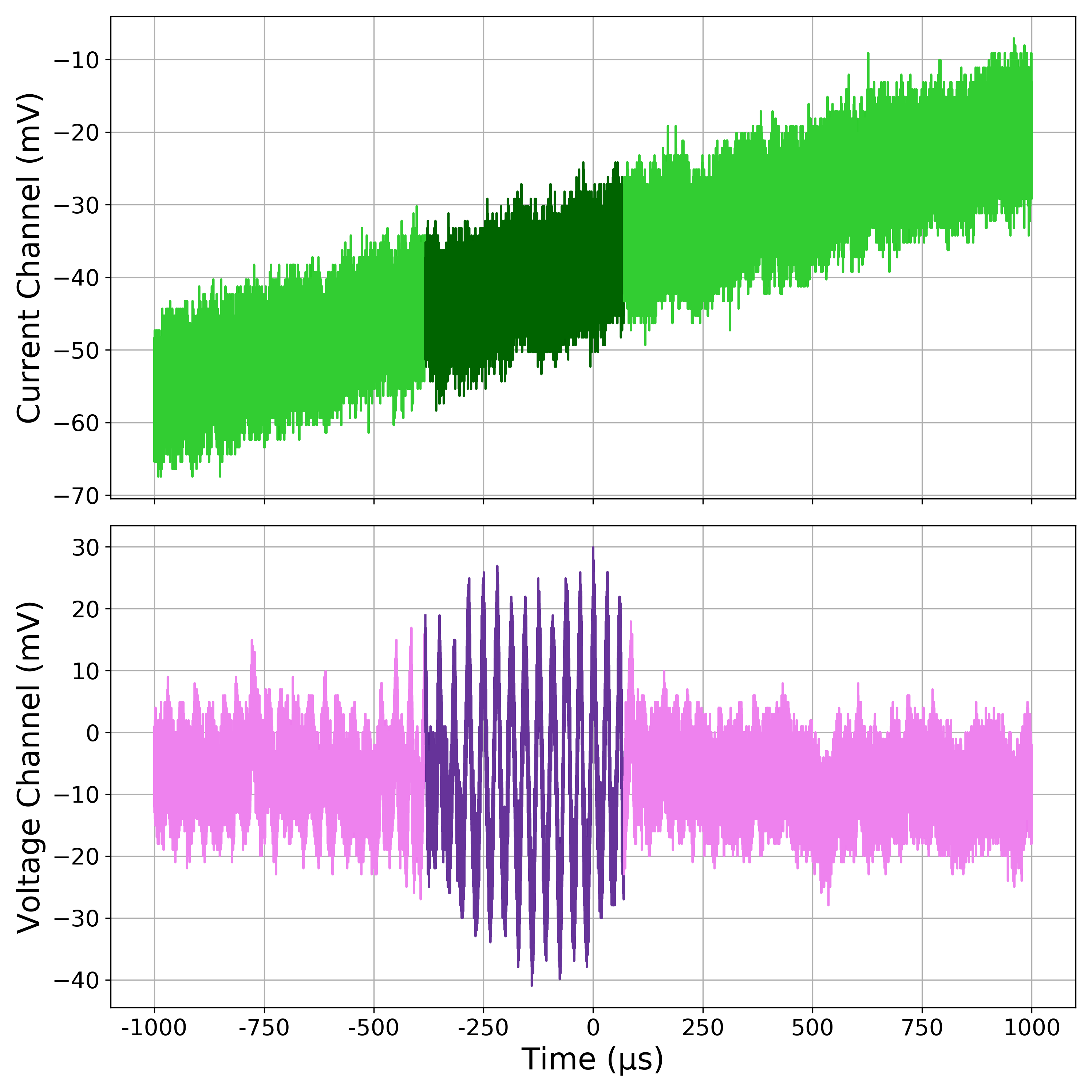}
    }
    \caption{Examples of the two types of non-radiation-induced perturbations. 
    Only voltage (pink) is affected; the current (green) channel behaves normally.}
    \label{fig_wrong_faults}
\end{figure}

\subsubsection{E3}
The experiment with the monoenergetic gamma field shows a different approach, since the radiation source, namely $^{60}$Co, is a radioactive material. 
A key distinction lies in the method of exposure characterization. Unlike NILE and ChipIR, which typically rely on direct flux measurements, the CALLIOPE irradiation chamber is mapped with dosimetry values. This dosimetry is performed before the actual experiment: a dosimeter \cite{calliope} is positioned inside the SQUID dewar filled with water to simulate the experimental environment during irradiation. With this method, we identify three spots in the irradiation chamber where to test the SQUID under different conditions.

At CALLIOPE, communication with the control room is provided through a long tunnel. The SQUID is connected to the control device in a concrete-protected area inside the irradiation chamber via a $\SI{7}{m}$ serial cable. From there to the oscilloscope, in the control room, the communication is established through two $\SI{30}{m}$ coaxial cables.   

The first run is performed with a dose rate of $\SI{8.84}{Gy/h}$, lasting for $\sim \SI{71}{minutes}$ and therefore implying a total absorbed dose of $\SI{10.59}{Gy}$. The second test shows a higher dose rate, $\SI{10.06}{Gy/h}$, and, despite a shorter exposure ($\sim \SI{38}{minutes}$), results in a higher total dose absorbed: $\SI{12.27}{Gy}$. The third test is performed close to the source to unequivocally determine the dose rate dependence on SQUID single event upsets. The dose rate tested is $ \sim \SI{233.4}{Gy/h}$, with an exposure time of $\SI{12}{minutes}$, for a total absorbed dose of $\SI{46.68}{Gy}$. To conclude, the fourth and fifth runs are performed again with a dose rate of $\SI{10.06}{Gy/h}$, by placing the SQUID in two different spots in the room in order to tentatively collect information on environmental noise. The exposure times are $\sim \SI{105}{minutes}$ and $\sim \SI{103}{minutes}$, with total doses of $\SI{33.26}{Gy}$ and $\SI{32.76}{Gy}$.

In total, at the CALLIOPE facility, the SQUID absorbed a dose of $\sim \SI{135.56}{Gy}$.

\subsubsection{E4-E5}

With the fourth experiment, we recreate E2 conditions to show the behavior of the SQUID under similar conditions after the exposure to the gamma field at CALLIOPE. The fifth experiment has the same aim, namely to replicate E1's conditions as closely as possible.
The positions in the respective rooms, the neutron fields on the device, and the experimental setups are reproduced analogously to E1-E2. Despite that, as explained in Section~\ref{sec:experimentalResults}, there was a need to test additional setups as well. Specifically, cables are changed to shorter ones (serial $\leq \SI{5}{m}$, coax $\sim \SI{1}{m}$), and the measuring setup is moved inside the irradiation rooms and shielded with borated polyethylene blocks.

\begin{figure}[t]
    \centering
    \includegraphics[width=1\columnwidth]{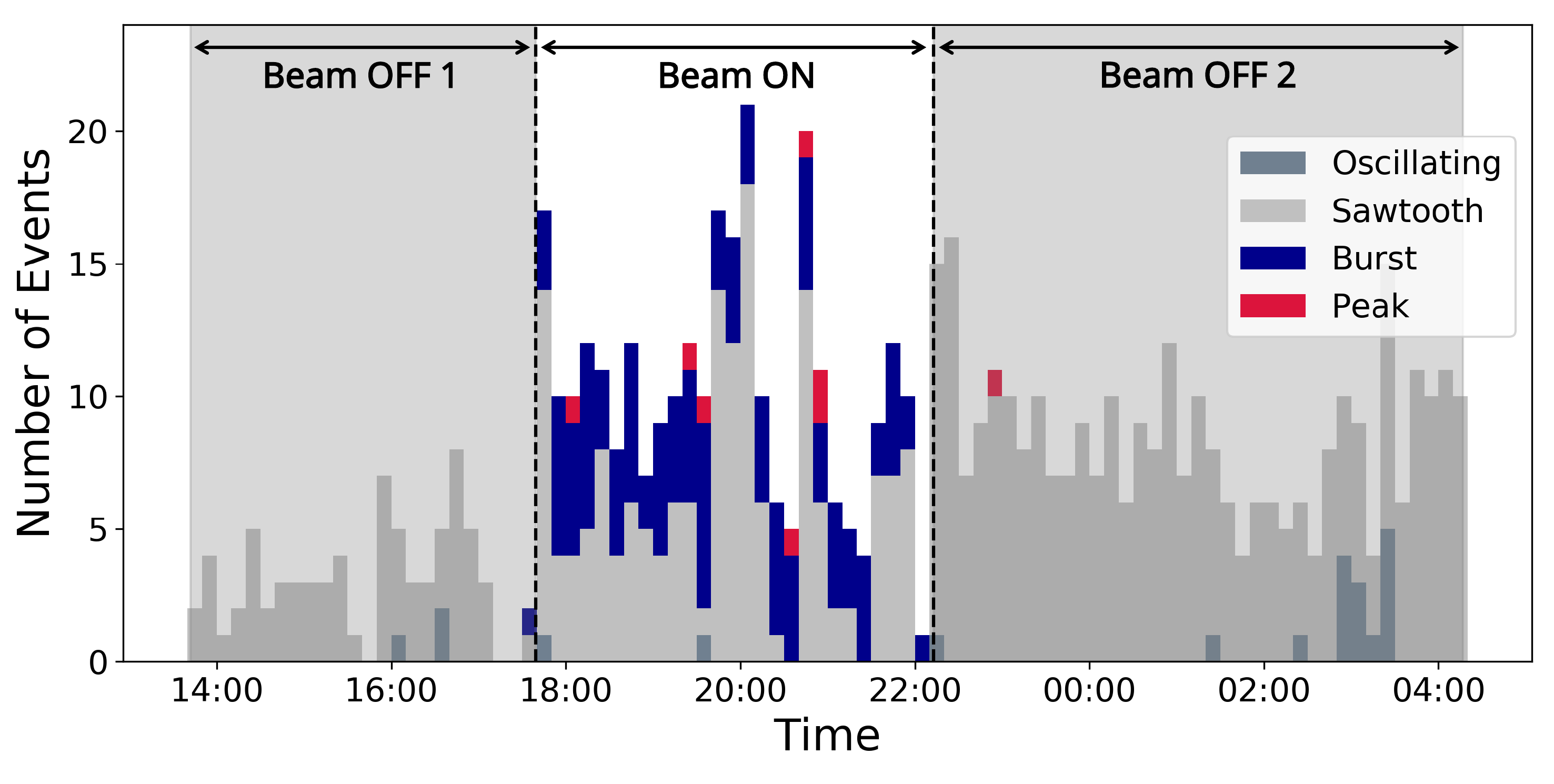}
    \caption{Counts during beam off (shaded areas) and beam om time at NILE. In gray, the non-radiation induced faults. In blue the burst type and in red the peak type faults.}
    \label{fig_timeline_NILE}
\end{figure}

\subsection{Separating Radiation- and Non-Radiation-Induced Events}
\label{subsec_non-radiation-induced}

The analog nature of the SQUID and its sensitivity to magnetic field fluctuations make its output intrinsically noisy. The probe in the dewar is shielded with a cylinder of $\mu$-metal and aluminum foil to attenuate any electromagnetic noise, however the occasional recordings of stochastic fluctuations are to be expected.

We set a trigger on the SQUID voltage channel and tune it at the optimal value required in order to maximize sensitivity to radiation events, while minimizing the recording of spurious effects. The trigger value is highly dependent on the environmental conditions and on the setup (specifically the length of the cables, which influences resistivity and capacitance). This trigger value is in the order of $O(\SI{10}{meV})$, specifically for experiments E1 and E2 it was set to $V_t=\SI{30}{mV}$ and $V_t=\SI{20}{mV}$ (represented as lines in Figure \ref{fig_normal_output}), respectively.



\begin{figure*}[!ht]
\centering
\begin{minipage}{.49\textwidth}
    \centering
        \includegraphics[width=.95\textwidth]{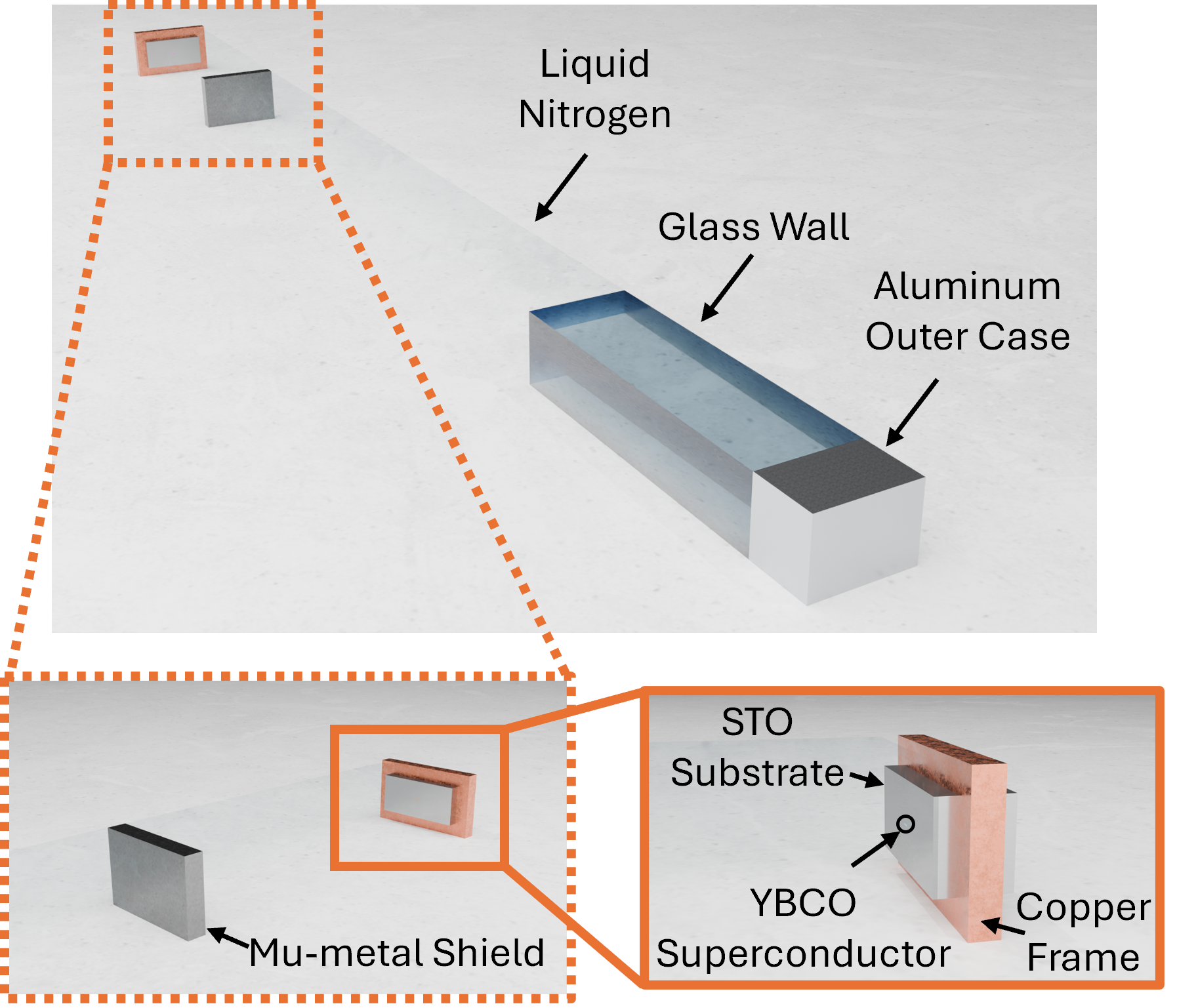}
        \caption{Geant4 simulation setup. A coring of the experimental setup is represented. The particles are fired facing the outer aluminum layer and must traverse multiple sources of shielding before reaching the SQUID. From outside: $\SI{2}{mm}$ of aluminum, $\SI{10}{mm}$ of glass, $\SI{29.5}{mm}$ of liquid nitrogen, $\SI{0.5}{mm}$ of mu-metal, and again $\SI{10}{mm}$ of Liquid nitrogen.}
        \label{fig:simulation_setup}
\end{minipage}%
\begin{minipage}{0.02\textwidth}
{
\quad
}
\end{minipage}%
\begin{minipage}{0.49\textwidth}
    \centering
        \includegraphics[width=.95\textwidth]{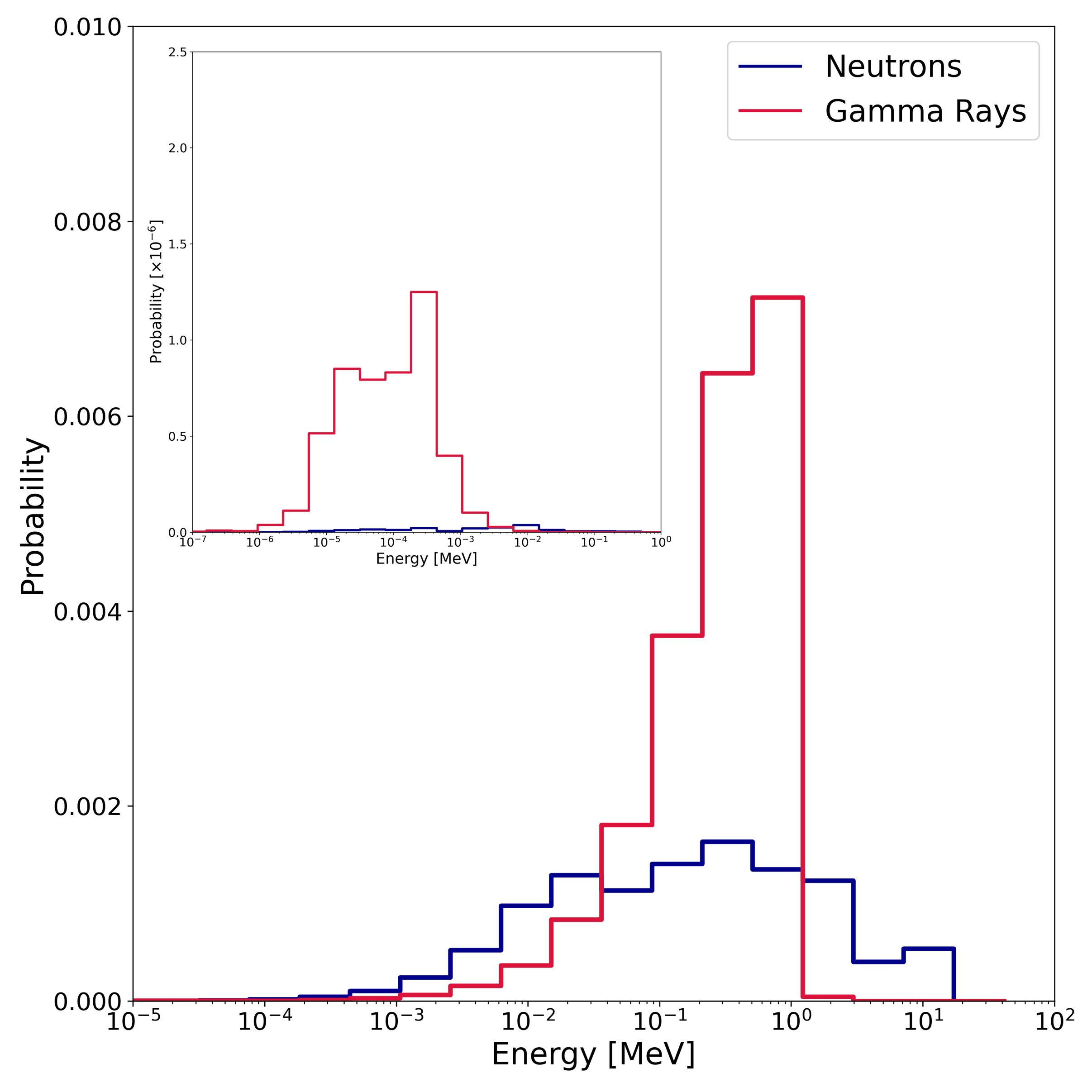}
        \caption{Energy deposition spectra on the SQUID for a test with $10^9$ particles. Neutrons are in blue, and gamma rays are in red. In the inner plot, a detail of the energy deposition on the superconductor.}
        \label{fig:charge_deposition_distribution}
\end{minipage}
\end{figure*}
Any event crossing the trigger value is saved by recording a time window of $\SI{2}{ms}$ centered around it. In post-processing, we investigate the events collected and differentiate between radiation-induced and non-radiation-induced events.
In Figure~\ref{fig_wrong_faults}, we show the two prototypes of spurious effects detected and classified as non-radiation-induced events. These events are therefore discarded from further analyses. This firm differentiation method that we propose is part of our contribution and is supported by four separate metrics. Conversely to the radiation-induced faults (depicted in Figure \ref{fig_bp}), the non-radiation events (i) barely cross the trigger value, (ii) only affect the voltage channel leaving the current unperturbed, (iii) affect the voltage channel following specific alteration shapes (single sawtooth-shaped and oscillation alterations), incompatible with the radiation-induced ones, and (iv) have an occurrence rate which is not correlated to the activity of the beam.
Given (i) and (ii), we suspect that most of the non-radiation-induced events are caused by a temporary modification of the I-V relation, probably induced by the SQUID leaving the superconducting regime. Indeed, outside of this region, any variation of the current (which is driven along a triangular signal) induces an ohmic variation of the voltage. The net behavior can be described by accounting for a transient modification of the critical current. 

To confirm our distinction strategy, in Figure~\ref{fig_timeline_NILE} we show the temporal distribution of the detected events of experiment E1, where we alternate \textit{beam OFF} and \textit{beam ON} periods. The occurrence rate of the radiation-induced events (red and blue) is correlated with the neutron irradiation. On the contrary, the sawtooth-shaped events (gray) happen with a variable occurrence rate ($r_s \in [0.3,0.9]$) throughout the entire experimental session, regardless of the neutron activity. Oscillating alterations (dark gray) are deemed not statistically significant, given that only a few of such events are recorded.

\begin{figure*}[t]
    \centering
    \subfigure{
        \includegraphics[width=0.48\textwidth]{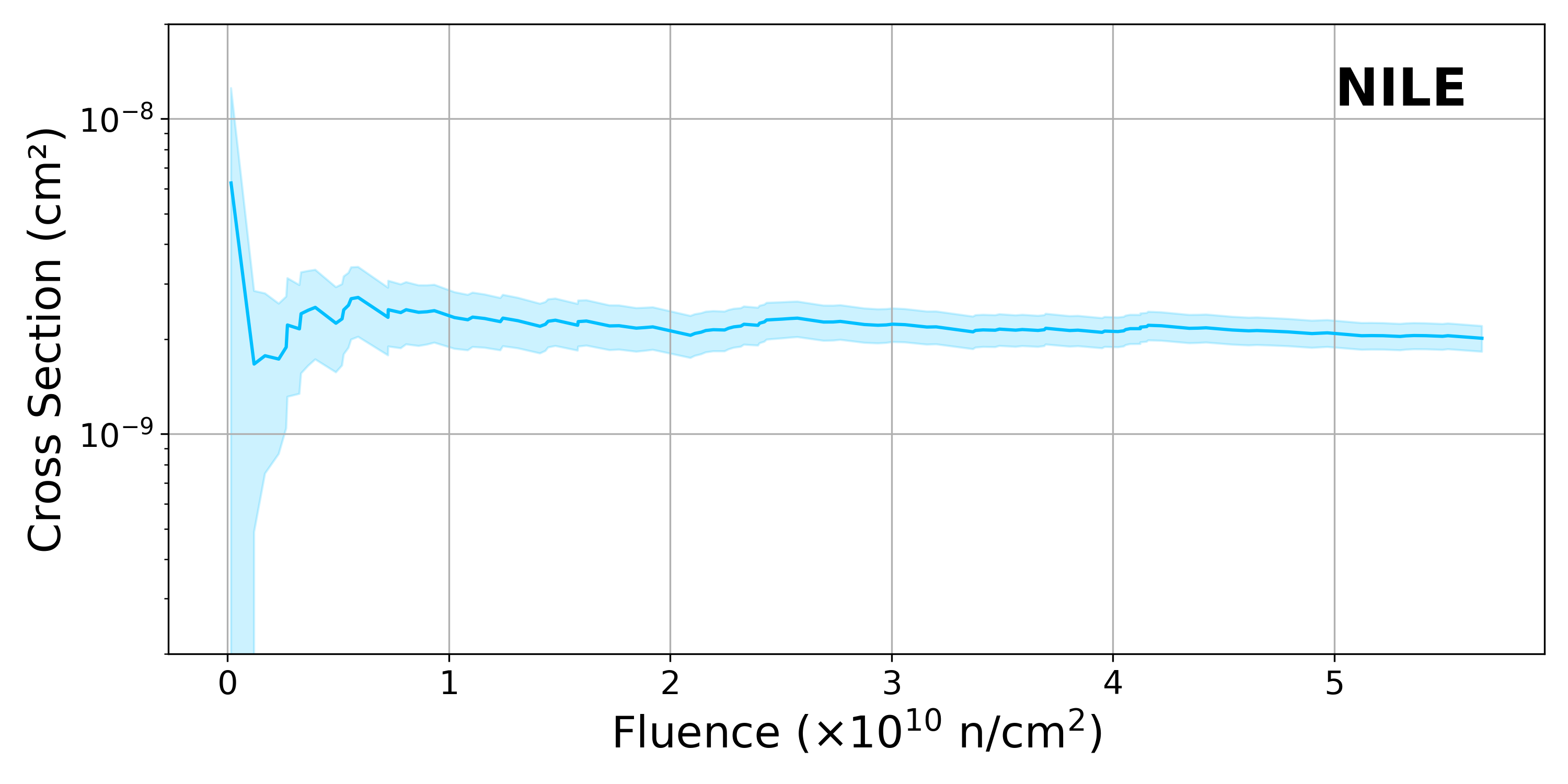}
    }
    \subfigure{
        \includegraphics[width=0.48\textwidth]{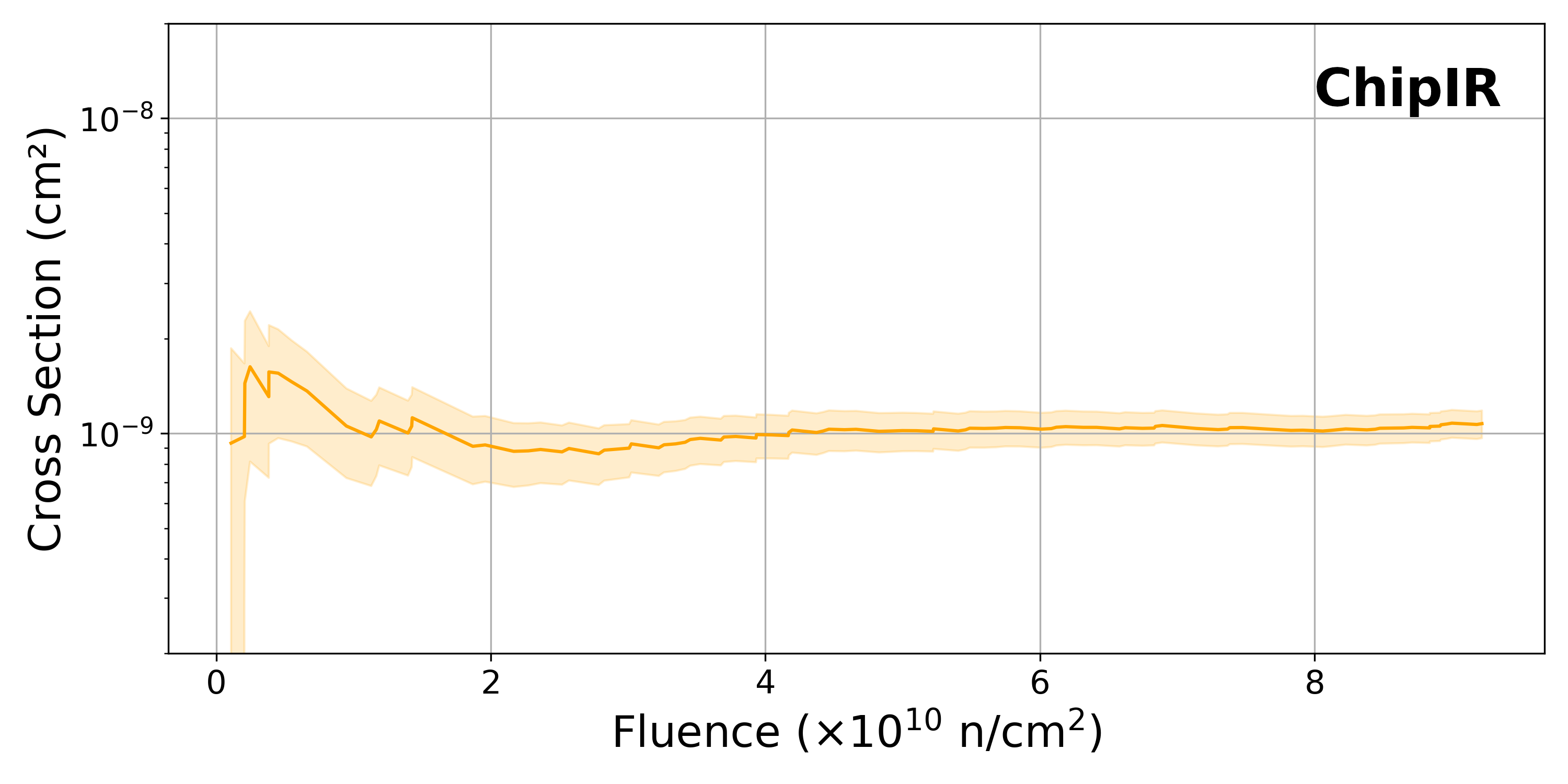}
    }
    \caption{Cross section at NILE (left, light blue) and ChipIR (right, orange) as a function of the neutron fluence.
The shaded area represents the statistical error, considering the $95\%$ confidence interval of a Poisson distribution.}
    \label{fig_cross_section}
\end{figure*}

\section{Simulations}
\label{sec:simulations}
To provide the full picture of the reliability problem on fundamental quantum devices, experimental observations are not sufficient. While the experiments focus on the response of the SQUID, to get a complete understanding of the phenomenon one needs to also measure the energy deposition details. Moreover,
there are very few reliability studies on fundamental quantum devices, and simulations are a crucial validation marker for our experiments.

\subsection{Simulation Environment}\label{subsec_simulationenv}
Simulations are carried out using Geant4, which is a Monte Carlo simulator for particle physics \cite{Geant4}. In Geant4, interaction mechanisms and physical processes are defined through physics lists. We develop a custom class based on \textit{QGSP\_BIC\_HP} physics list, which resolves both ionizing, indirect ionizing (for neutral particles), and non-ionizing processes. To these built-in Geant4 features, we add the resources of G4CMP toolkit \cite{g4cmp}, which brings to the picture low-energy physics processes. Specifically, after the computation of the dynamics of the energy deposition in a material, G4CMP accounts for the propagation of energy by different means. The first step is the creation of $e^-/h^+$ pairs, which then either recombine, generate phonons, or emit Neganov-Luke phonons. Phonons can downconvert to less energetic phonons (ballistic) and propagate, undergoing reflections and refractions on the surfaces.

\subsection{Simulation Setup}
Geant4 offers a fine code-based design functionality where we model the device presented in Section~\ref{subsec_SQUID}. In addition to the mentioned components, we place a fictional Copper frame around the STO substrate, covering $22 \%$ of the substrate external sources and accounting for the acoustic coupling of the SQUID with the environment. In Figure~\ref{fig:simulation_setup}, we show the geometric structure within Geant4. Notably, apart from the SQUID itself, the whole experimental setup is reconstructed: the liquid nitrogen filling, the glass container, and the aluminum shielding. For Geant4 energy deposition simulations, the SQUID is modelled with the same materials, and energy deposition is calculated according to the specific structure and properties of each component. Conversely, the energy propagation computed with G4CMP is limited to a handful of elements, and STO is not among them. We then perform the simulation with a similar lattice structure and only consider relative results coming from this simulation branch.    

Neutrons and gamma rays are fired from a rectangular surface as large as the substrate, placed $\SI{10}{nm}$ away from the external aluminum layer. Particles are generated in empty space to reduce numerical instability. As such, they are fired perpendicularly to the SQUID, and immediately enter the outermost shield of the dewar. 

\subsection{Simulation Runs}
We perform two distinct simulation experiments, with different purposes. In the first case, we run a Geant4 only simulation, involving $10^9$ $\SI{14}{MeV}$ neutrons and $10^9$ $\SI{1.25}{MeV}$ gamma rays, where only the deposition spectrum is measured. In the second case, we add the G4CMP toolkit and perform $\SI{5e5}{}$  injections for each particle species and collect data on the energy propagation. Specifically, we measure when secondary particles are created, where and when they are annihilated, and how many of them there are. The number of simulated particles is reduced to because of the larger computational requirements of such additional physical processes.
\section{Results}
\label{sec:experimentalResults}

As discussed in Section~\ref{subsec_non-radiation-induced}, an important finding of this research is the ability to separate radiation-induced faults from the background effects recorded by the measuring setup. Experimentally, the latter evidently stand out by showing a larger signal amplitude, by affecting the signal following specific patterns, and concerning both the current and voltage channels.
Considering the corruption mechanism presented in Section~\ref{sec:background}, the last observation is well supported by theory, since the breaking of Cooper pairs should induce a current change, and consequently a voltage rise.

Following this distinction of the radiation cause of faults, we proceed with their counting and classification. For this purpose, we take into account the runs from experiments E1 and E2, which are conducted close to one another in time and involving two different mixed neutron-gamma fields, as described in Section~\ref{sec:experimentalMethodology}.

\begin{figure*}[t]
    \centering    \includegraphics[width=2\columnwidth]{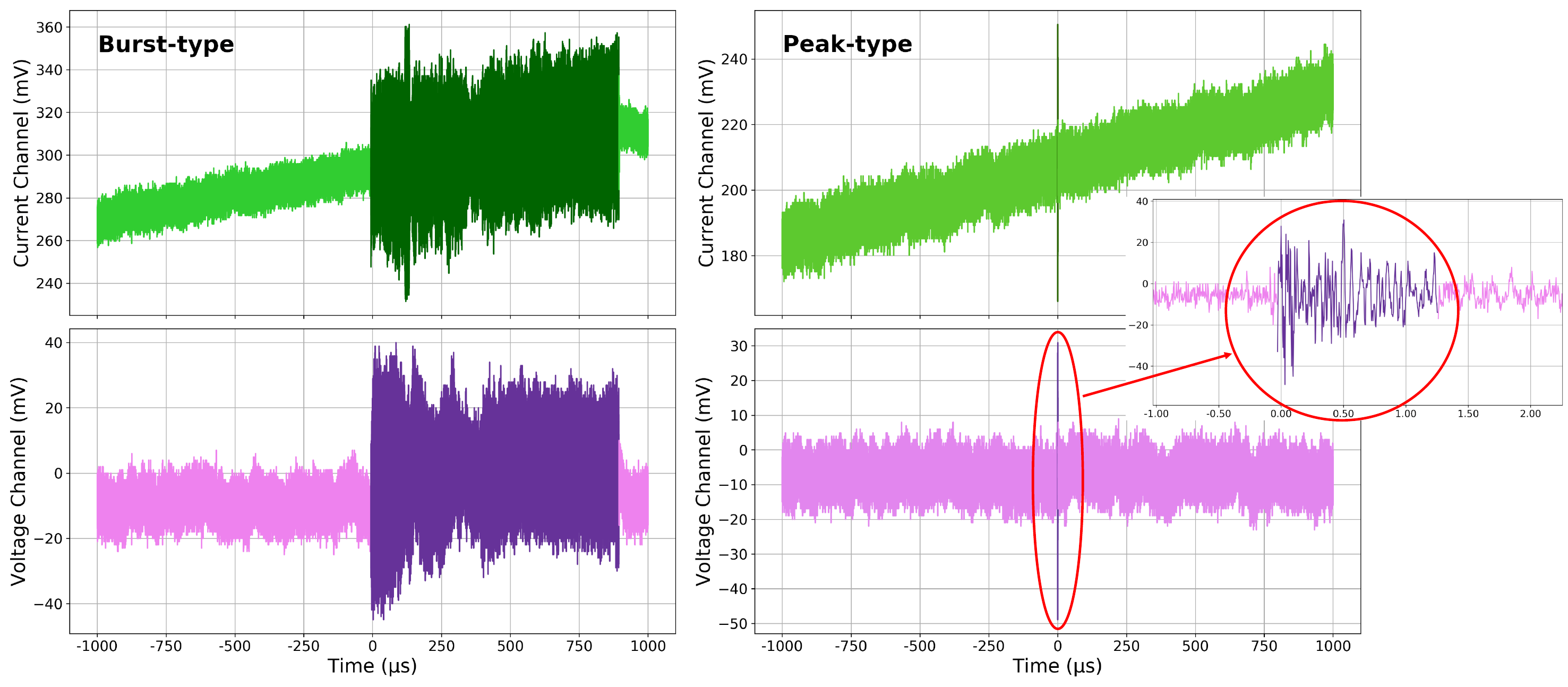}
    \caption{On the left, an example of a burst-type fault. Both the current and voltage channels are corrupted for $\sim\SI{900}{\mu s}$. The maximum amplitude is $\sim \SI{90}{mV}$. On the right, a prototype of peak-type fault. Both channels are corrupted with a maximal amplitude similar to the burst-type case. Conversely, the duration of the event is $\sim\SI{1.2}{\mu s}$. In the insight picture, 
    a zoom-in of the voltage channel  (the current channel is analogous).}
    \label{fig_bp}
\end{figure*}

In Figure~\ref{fig_cross_section}, we show the cross section for NILE (E1) and ChipIR (E2).
The plots are generated by computing the cross section ($\sigma_n \doteq$ number of faults / fluence from neutron counts) after each recorded event iteratively and accounting for the respective statistical error, considering the $95\%$ confidence interval of a Poisson distribution. The plots also show that the cross section does not increase as the neutron fluence increases, i.e., the effects on the SQUID are transient, without any dose dependence. The SQUID-neutron cross section is $\sigma_n^{\text{NILE}}=2.0 \pm 0.2 \times 10^{-9} \text{cm}^2$ at NILE and $\sigma_n^{\text{ChipIR}}=1.1 \pm 0.1 \times 10^{-9} \text{cm}^2$ at ChipIR. Since the facilities are built for chip testing, no real-time gamma ray counting is implemented. However, following the calculations from legacy experiments and simulation presented in Section~\ref{subsec_radiationfacilities}, it is at least possible to conjecture a cross section value including the secondary gamma ray field for the experiment at NILE: $\tilde{\sigma}_{n \gamma}^{\text{NILE}} \sim 1.87 \times 10^{-9} \text{cm}^2$. 

The cross section values at NILE and ChipIR naturally differ. Indeed, at NILE the neutron energy spectrum is 14 MeV, whereas at ChipIR it emulates the atmospheric spectrum. Our previous simulations~\cite{simulazioni} showed that the interaction probability and deposition are considerably affected by the neutrons' energy (1 MeV neutrons have a 1/10 probability to interact with the SQUID compared to 10 MeV and 100 MeV neutrons). As explained in Section~\ref{subsec_experimentdetails}, the cross section at ChipIR is measured following the electronics standard, so considering the counts of neutrons $>\SI{10}{MeV}$. The sensitivity of superconducting-based quantum technologies to low-energy neutrons (despite being $1/10$ compared to high-energy neutron) is then not considered, inducing an overestimation of the value reported in Figure~\ref{fig_cross_section}. Given the importance of low-energy neutrons for quantum devices, a neutron counter with a lower (at most $\sim \SI{1}{MeV}$) energy threshold is needed for more accurate measurements. Together with the instrumental scientist, we plan to enable the facility neutron counter to include $<\SI{10}{MeV}$ neutrons at ChipIR.

The first part of the analysis considered all the events recorded and identified as radiation-induced. Hereinafter, to go beyond the mere count of electrical faults, we propose a categorization of the events recorded based on the specific values collected by the current and voltage channels of the SQUID.
For every fault, we collect $2 \times 10^7$ data points over $\SI{2}{ms}$: from 1 ms before to 1 ms after the triggering event (set to be at $t \doteq 0$). The first quarter of the saved interval $[\SI{-1}{ms},\SI{-0.5}{ms}]$ is used to define the average behavior of the channels in a non-corrupted state (we name it the reference window). Then in the second quarter, we seek the manifestation of the fault. Indeed, although the triggering event always occurs at $t=\SI{0}{s}$, a gradually increasing tail is likely to appear before the threshold value is actually crossed. To record the end of the event, we look for the first window of $\SI{80}{\mu s}$ in the last half of the recorded data in which the signal is once again compatible with the behavior in the reference window. Besides considering the time length, in the categorization process, we take into account the fault amplitude by looking at the local variation of the voltage channel in rolling windows of $\SI{100}{ns}$. 


Analyzing the waveforms recorded during the experiment, we classify the radiation effects into two categories: \textit{burst-type} and \textit{peak-type}. The former is characterized by a long perturbation of current and voltage, while the latter is a short variation of the two channels. 
Figure~\ref{fig_bp} (left) shows a prototypical example of a \textbf{burst-type} fault. Burst-type events have a perturbation length in the order of $ \sim \SI{100}{\mu s}$. The events collected in this category show many different shapes, according to the frequency at which the signal is disturbed. We observe bursts where all the data points are corrupted and some others where the signal is altered every few \text{$ns$} to \text{$\mu s$}. Finally, some events last longer than the recorded window; in these cases, the fault duration is undetermined. Interestingly, the amplitude of the perturbation, as well as the time length, is homogeneous over all burst-type faults (between $\SI{50}{mV}$ and $\SI{250}{mV}$), supporting the classification into a single category.

We name the second fault category \textbf{peak-type}, represented in Figure~\ref{fig_bp} (right). These events are much shorter compared to burst-type, in the order of $\SI{100}{ns}$. The amplitude is similar to the burst one, between $\SI{50}{mV}$ and $\SI{200}{mV}$, with a couple of outliers around $\SI{400}{mV}$. The shape of these events, represented in the insight of Figure~\ref{fig_bp}, is the same for every fault. 

\begin{figure}[t]
    \centering
    \includegraphics[width=1\columnwidth]{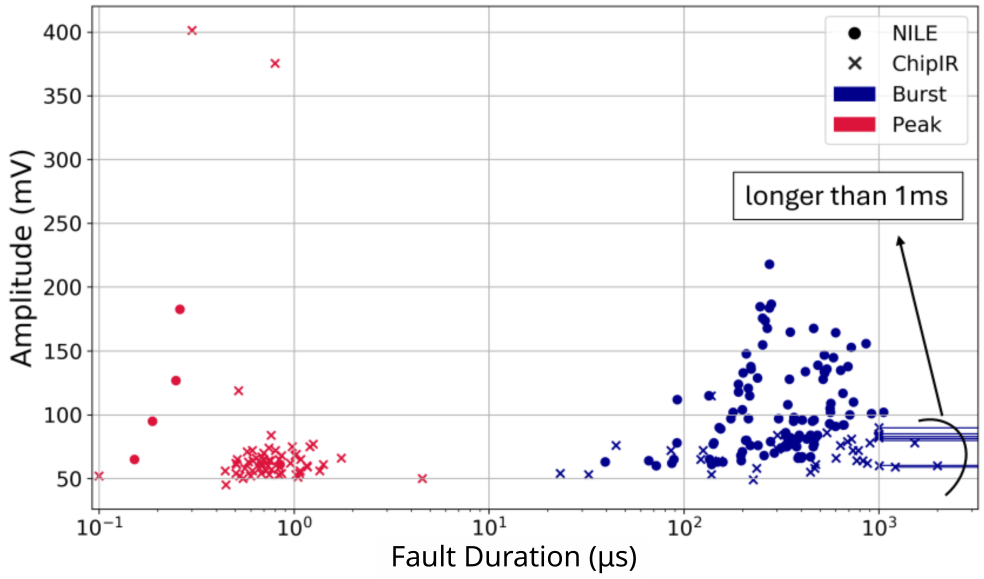}
    \caption{Fault classification by amplitude and fault length. Peak-type (red) and burst-type (blue) faults observed during E1 at NILE (dots) and during E2 at ChipIR (crosses). Some bursts (lines dragged to the right) at ChipIR were longer than 1 ms.}
    \label{fig_error_classification}
\end{figure}

\begin{figure}[t]
    \centering
    \includegraphics[width=1\columnwidth]{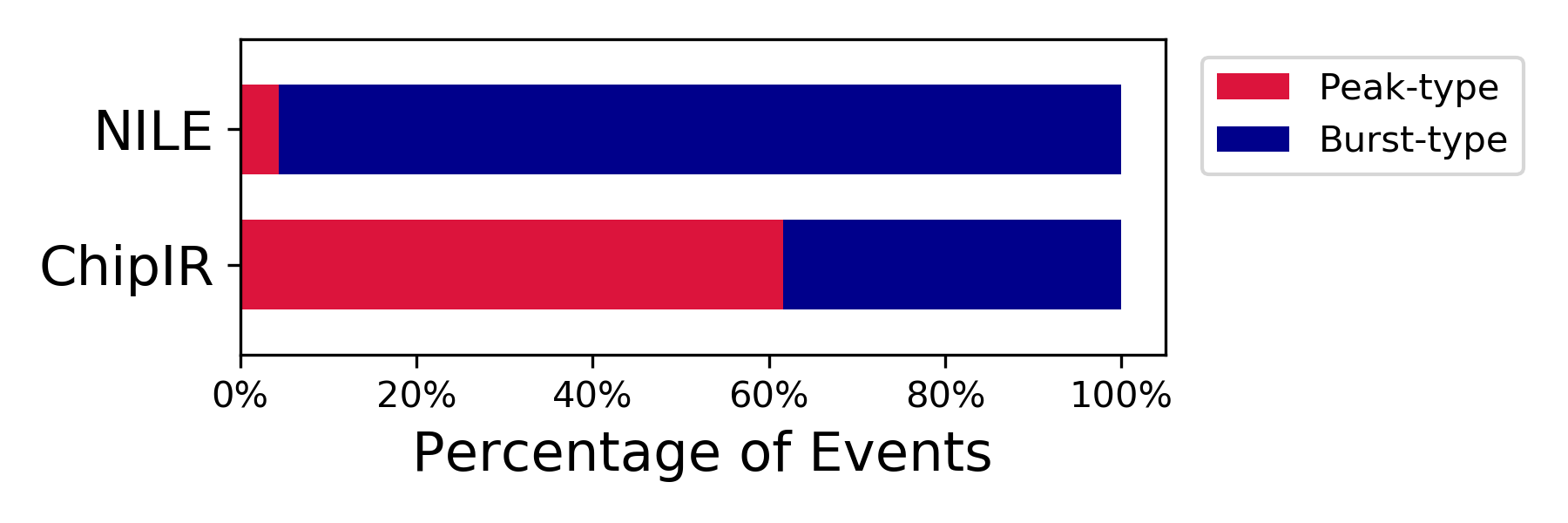}
    \caption{Percentage of observed peak-type (read) and burst-type (blue) events for experiment E1 at NILE and experiment E2 at ChipIR.} 
    \label{fig_cs_classification}
\end{figure}

The classification of the radiation faults is not merely a matter of nomenclature; it involves a deeper meaning related to the characteristics of the irradiated particles. In Figure~\ref{fig_error_classification}, we present a diagram collecting all the events recorded among NILE (dots) and ChipIR (crosses). In red are the peak-type faults, and in blue are the burst-type faults. The two categories are well separated according to their time length (Fisher Discriminant Ratio: 1.98). Peaks and bursts recorded in the two experiments show similar characteristics, becoming indistinguishable. Conversely, the relative ratio between the two fault types differs. At NILE (dots), the peak-type faults recorded are rare; on the contrary, at ChipIR they represent $60\%$ of the total events. This information is highlighted in Figure~\ref{fig_cs_classification}, where we show explicitly the peak-to-burst ratio in the two experiments.

As we explained in \ref{subsec_radiationfacilities}, the gamma ray abundance is very different between the two experiments ($7\%$ of neutron flux at NILE, undefined, but certainly higher, at ChipIR). Therefore, Figure~\ref{fig_cs_classification} suggests a correlation between the fault-type and the particle causing it. It was to verify this hypothesis that we performed experiment E3 with only a gamma ray field. Surprisingly, the result after a total irradiation of $\SI{13.5}{krad}$ is clear: no faults are recorded when the SQUID is exposed only to a gamma ray field.
The result is unexpected, considering the insight from the simulations. Geant4 computations suggest that the energy deposition spectrum between neutrons at $\SI{14}{MeV}$, as at NILE (E1), and gamma rays at $\SI{1.25}{MeV}$, as at CALLIOPE (E3), is different, but with an interaction probability higher for the latter species of particles. On the other side, Figure~\ref{fig:charge_deposition_distribution} shows that in $\sim 1/4$ of the neutron interactions, the energy released in the device is greater than the initial energy of the gamma rays tested at CALLIOPE. To investigate the role of energy on the device response, we employ G4CMP simulations for the energy propagation. For the limitations of the models implemented in the toolkit, the absolute results of these simulations can not be taken into account. However, it is valid and valuable to compare the neutron and gamma simulations of the charge and phonon propagation. The different energy deposition spectra consequently reflect on the secondaries' dynamics within the substrate. Specifically, for equal fluence, the energy absorbed by the substrate when firing neutrons is $1.33 \pm 0.05$ times greater than the gamma ray setting. This reference is attenuated when we consider the ratio between the number of phonons absorbed by the superconducting layer in the two experiments: $1.10$. We also look to the number of initial particles that are actually generating secondaries, which are absorbed by the superconducting layer, a metric intrinsically related to the experimental cross section presented above. Here, the ratio between the neutron and the gamma ray experiments data is $0.41$. This result actually predicts a higher interaction rate in experiment E3 at CALLIOPE.
Ultimately, this suggests that experiment E3 is unsuccessful, and it is likely that in a different environment, gamma rays affect superconducting quantum devices. The main hypothesis behind the failure of the experiment is attributed to the longest cable configuration among the five tests. This is potentially attenuating the signal measured by the SQUID to such a level that it becomes indistinguishable from the environmental noise. Due to the layout of CALLIOPE irradiation chamber, no other setup design is possible without exposing the control device to an unknown gamma ray field. Since the amount of energy deposited by gamma rays is less compared to neutrons, another option could be that the gamma-ray-induced faults are too quick to be recorded with our experimental setup. About this conjecture, we performed another comparative simulation to analyse the persistence of secondary particles induced by neutrons and gamma rays. In particular, we measured the times when phonons and $e^-/h^+$ are absorbed by the superconductor. Considering the $99.5\text{th}$ percentile, simulations show that the dissipation of gamma rays is indeed quicker, but just by a factor $1.07$.  The absorption time is then practically equal, and, considering the sampling interval of $\SI{4}{ns}$ of our setup, the hypothesis of an effect too quick to be observed is most likely wrong.

As for classical technologies, while dealing with gamma rays, SEUs are not the only concern; it is likely that gamma rays damage CMOS at an atomic level as a consequence of the dose absorbed. Indeed, one month after experiment E3, we performed experiments E4 and two weeks after E5, which aimed to reproduce E2 and E1 as closely as possible. The results of E4 and E5 clearly show a modification of the device under test. Specifically, the SQUID becomes extremely sensitive to any electromagnetic signal. With a shielding analogue to the one used in E1 and E2, the sole presence of an electrical circuit close to the probe affects the output signal by orders of magnitude. We employ different shielding structures and add custom disturbance elements in order to characterize the new sensitivity of the device. However, the only result is a beam-uncorrelated recording of events impossible to interpret. Some of the shapes of the output channels vaguely recall the ones in Figure~\ref{fig_wrong_faults} and Figure~\ref{fig_bp}, but with different amplitudes and time lengths. It is evident that the events recorded are not radiation induced, and likely that the difference observed is induced by the severe exposure to gamma rays.


\section{Conclusions}
\label{sec:conclusions}

The results of this work are tightly packed with information, reflecting the challenging nature of both the experiments and the subject at hand. While we observed critical radiation-induced effects on superconducting quantum devices, we also identified limitations in the experimental setup and the facilities used to detect these events.

Experiments E1 and E2, where the SQUID was exposed to a mixed neutro-gamma ray field, undoubtedly showed the device's sensitivity to external radiation. Moreover, they showed the feasibility of tests with quantum technologies in the facilities of ISIS Muon and Neutron Source. The data collected during these experiments helped us in characterizing the SQUID response, classifying faults according to their features into \textit{bursts} and \textit{peaks}.
The conjecture of the correlation between the nature of the primary particle and the induced fault type has been investigated. However, experiment E3 cast a shadow on this hypothesis, since the exposure of the device to gamma rays resulted in zero faults being detected. Nonetheless, Geant4 simulations firmly indicate a sensitivity of the SQUID to gamma rays and neutrons, thus suggesting that E3 could have suffered from issues in the setup. Indeed, the requirement of a considerably longer (compared to the other setups) serial cable and of the long coaxial cables in E3's setup might have led to systematic energy losses of the SQUID signal. A possible solution would be to redesign the setup, including a signal amplifier between the SQUID and the serial cable. Despite not having detected any fault when testing the pure field of gamma rays, the exposure of the SQUID to the latter was not without consequences. The following experiments, E4 and E5, showed such a stark increase in the device's sensitivity to external stimuli that even basic measurements became hard to impossible.

Unsurprisingly, the issue of noise is the main concern for tests involving superconducting quantum devices. The absence of spurious electromagnetic background fields inside the irradiation rooms, or at least the knowledge of their precise intensity and spectrum, is a crucial characteristic for facilities aiming to test quantum technologies going forward. Additional precautions, such as the shielding cables and devices, can be employed at the setup level; however, one must resort to these only as mitigation measures, and not as primary solutions.

Finally, in the context of experiments involving high-temperature superconductors, we believe that making use of a cryostat designed for irradiation studies that go to $\SI{0}{K}$ could be beneficial for the experiments. This would considerably reduce the thermal noise responsible for the uncertainty on the SQUID measurements, making even smaller alterations detectable. For our future experiments with SQUIDs, we thus aim to use a different a higher higher-performance cooling system.

\appendices

\section*{Acknowledgments}
This work was possible thanks to the collaboration of many people. The test at CALLIOPE has been possible thanks to Alessia Cemmi, Ilaria Disarcina, Beatrice D'Orsi, Giuseppe Ferrara, Jessica Scifo, and Adriano Verna. The experimental equipment and the preliminary tests were available thanks to Emanuela Callone, Mirko Lobino, and Lucio Pancheri. We finally thank Robin Cantor and Nicolò Crescini for the support with the SQUID.

\clearpage

\bibliographystyle{IEEEtran}
\bibliography{refs}

\end{document}